%% file: PoincareMZ0.tex
\input Definitions.tex

\centerline{\bf SEMI-CLASSICAL RESONANCES}

\centerline{\bf ASSOCIATED WITH A PERIODIC ORBIT OF HYPERBOLIC TYPE}
\bigskip
\centerline{Hanen LOUATI ${}^{1,2}$, Michel ROULEUX ${}^{2}$}
\medskip
\centerline {${}^{1}$ Universit\'e de Tunis El-Manar, D\'epartement de Math\'ematiques, 1091 Tunis, Tunisia}

\centerline {e-mail: louatihanen42@yahoo.fr}

\centerline {${}^{2}$ Aix Marseille Univ, Univ Toulon, CNRS, CPT, Marseille, France} 

\centerline {e-mail: rouleux@univ-tln.fr}               
\bigskip
\noindent {\it Abstract}: We consider in this Note resonances for a $h$-Pseudo-Differential Operator $H(x,hD_x;h)$ on $L^2(M)$
induced by a periodic orbit of hyperbolic type, as arises for Schr\"odinger operator with AC Stark effect 
when $M={\bf R}^n$, or the geodesic flow on an axially symmetric manifold $M$, extending Poincar\'e
example of Lagrangian systems with 2 degrees of freedom. 
We generalize the framework of [G\'eSj], in the sense that
we allow for hyperbolic and elliptic eigenvalues of Poincar\'e map,
and look for so-called semi-excited resonances with imaginary part of magnitude $-h\log h$, or $h^s$, with $0<s<1$.
\medskip
\noindent {\bf I. Hypotheses and the main result}.
\smallskip
For simplicity, we present our results with $M={\bf R}^n$. 
Let $H(x,hD_x;h)$ be a self-adjoint $h$-PDO on $L^2({\bf R}^n)$
$$H^w(y,hD_y; h)u(y;h)=(2\pi h)^{-n}\int\int e^{i(y-y')\eta'/h}H({y+y'\over2},\eta';h)u(y')\, dy'\, d\eta'\leqno(1.1)$$
We assume it has Weyl symbol $H(y,\eta;h)\in S^0(m)$, where $m$ is an order function (for example $m(y,\eta)= (1+|\eta|^2)^M$), and
$$S^N(m)=\{H\in C^\infty(T^*{\bf R}^n): \forall\alpha\in{\bf N}^{2n}, \exists C_\alpha>0,
|\partial^\alpha_{(y,\eta)}H(y,\eta;h)|\leq C_\alpha h^N m(y,\eta)\}$$
with the semi-classical expansion
$H(y,\eta; h) \sim H_0(y,\eta)+hH_1(y,\eta)+\cdots, h\rightarrow 0$.
Here $H_0$ is the principal symbol of $H$, $H_1$ its sub-principal symbol. We assume that
$H(y,\eta;h)$ is elliptic (i.e. $H+i$ elliptic) and defines an analytic symbol in the sense of [Sj]  in a sector
$$\Gamma_0=\{(y,\eta) \in T^*{\bf C}^n : |\im (y,\eta)|\leq\const \langle\re (y,\eta)\rangle\}$$ 
Let the energy surface $H_0^{-1}(E_0)$ be regular for some $E_0\in{\bf R}$, that we may set up to 0. So the Hamiltonian vector field
$X_{H_0}$ has no fixed point on $H_0^{-1}(0)$, hence on nearby energy surfaces $H_0^{-1}(0)$. 
Let $\Phi^t = \exp(tX_{H_0}) : T^\ast {\bf R}^n \rightarrow T^\ast {\bf R}^n$ and
$$K_E = \{\rho \in T^\ast {\bf R}^n, H_0(\rho) = E, \Phi^{t}(\rho) \ \hbox{doesn't grow to infinity} \ \hbox{as} \ |t|\to\infty\}\leqno(1.2)$$
be the trapped set at energy $E$. We assume that 
$K_0=\gamma_0$ is a periodic orbit of period $T_0$.
Let ${\cal P}_0$ be Poincar\'e map (first return map), acting on a Poincar\'e section $\Sigma(\rho)\subset T^*{\bf R}^n$, $\rho\in\gamma_0$.
Assume also that 1 is not an eigenvalue of $d{\cal P}_0|_{\gamma_0}$, then $\Sigma(\rho)$ is transverse to the center manifold 
$\overline{\gamma}$, identified with a neighborhood of the zero-section in $T^*{\bf S}^1$. 
Each $\Sigma(\rho)$, $\rho\in\gamma_0$, identifies with
$\Sigma\approx T^*{\bf R}^d$ (locally along $\overline{\gamma}$, modulo the action of Hamiltonian flow). 
Both $\overline{\gamma}$ and $\Sigma$ are symplectic manifolds, and for small $E$: 
$K_E=\gamma_E$ is a periodic orbit of period $T_E$, $\bigcup_E\gamma_E=\overline{\gamma}$. For $\rho\in\gamma_0$, let
$\lambda_j, 1\leq j\leq 2d=2(n-1)$ be the eigenvalues of $A_0(\rho)=d{\cal P}_0(\rho) : {\bf C}^{2d} \rightarrow {\bf C}^{2d}$
(Floquet multipliers). 
The space ${\bf C}^{2d}$ has the orthogonal symplectic decomposition in (generalized) eigenspaces $F_\lambda$
relative to the family $(\lambda_j)_{1\leq j\leq 2d}$. We are interested in the case where 
$A_0(\rho)$ is partially hyperbolic, i.e. has at least one eigenvalue $\lambda$ of modulus $\neq1$.
Assume also that Poincar\'e map is non degenerate, i.e. $F_{\pm1} = \{0\}$, and also $F_{\lambda} = \{0\}$ for all $\lambda \leq 0$.
We say that $\lambda\in {\bf C}$ is {\it elliptic} (ee for short) if
$|\lambda|=1$ ($\lambda \neq \pm 1$) and {\it hyperbolic} (he) if $|\lambda| \neq 1$; 
if moreover $\lambda\in{\bf R}$ we call it {\it real hyperbolic} (hr)
and {\it complex-hyperbolic} (hc) otherwise.
Under the last assumption we can define $B=\log A$. Eigenvalues 
$\mu=\mu(\lambda)=\log\lambda$ of $B$ (Floquet exponents) verify $\mu(\overline{\lambda})=\overline{\mu(\lambda)}$.
Accordingly, exponent $\mu$ is said ee if $\re\mu=0$, hr if $\mu\in{\bf R}\setminus0$, and
hc if $\mu\in{\bf C}\setminus{\bf R}$. 
So eigenvalues of $B$ have the form $\mu_j, -\mu_j,\overline{\mu_j},-\overline{\mu_j}\neq 0$, $\re \mu_j\geq0$,
with same multiplicity. Let $b(\rho) = {1\over2}\sigma(\rho, B\rho)$ (Hermitian form), and $r$ be the number of distinct $\mu_j$'s. 
For simplicity, assume $r=d$, hence $b$ diagonalizable. We know [Bry] that
in a suitable basis $b(\rho)$ is a linear combinaison of elementary quadratic polynomials $Q_j$. 
If $\mu_j\in i{\bf R}$ (elliptic sector), we choose in $F_{\lambda_{j}}\oplus F_{-\lambda_{j}}$ 
symplectic coordinates (``harmonic oscillator coordinates'') such that $Q_j={1\over2}(x_{j}^2 + \xi_{j}^2)$.
If $\mu_j\in{\bf R}$ (real hyperbolic sector), one has $Q_j=x_j\xi_j$, while
in the complex-hyperbolic sectors, where $\mu_j=c_j+id_j$, one has $Q_j(x,\xi)=c_jQ'_j(x,\xi)-d_jQ''_j(x,\xi)$, with
$Q'_j(x,\xi)=x_{2j-1}\xi_{2j-1}+x_{2j}\xi_{2j}$, $Q''_j(x,\xi)=x_{2j-1}\xi_{2j}+x_{2j}\xi_{2j-1}$.  
In suitable complex symplectic coordinates, $Q_j$ has always the form $Q_j=x_j\xi_j$.
The $Q_j$'s play an important r\^ole, since they are formally  ``transverse eigenvectors'' for $H$, microlocalized near $\gamma_0$.
\smallskip
Our next Hypothesis is relative to {\it partial hyperbolicity} of Poincar\'e map,
in the sense that there exists $j \in \{ 1,...,r\}$, such that $\re\mu_j>0$.
For hyperbolic dynamic systems, we know [A] that generically only one $\mu_j$ has $\re\mu_j>0$. 
Let $F_{\mu_{j}}$, $\re \mu_j\geq0$ denote again the eigenspace associated with $\mu_j$. 
We can rewrite the decomposition of
${\bf C}^{2d}$ in the sum of unstable space $F^+$ and stable space $F^-$ :
$$F^+ = \bigoplus_{j=1} ^{r=d} F_{\mu_{j}}, \quad F^- = \bigoplus_{j=1} ^{r=d} F_{-\mu_{j}}$$
where $F^\pm \simeq {\bf C}^d$ are (complex) Lagrangian susbpaces of ${\bf C}^{2d}$ 
invariant under the flow of $X_b$, and such that if 
there exists an elliptic element ($\re\mu_k=0$), then for small $\theta>0, 
e^{-i\theta} X_b$ is ``expansive'' on $F^+$, ``contractive" on $F^-$. We call the elliptic element with positive imaginary part
an ``eigenvalue of the first kind''. 
Elliptic elements contribute to the center manifold. So generally we need consider 
Hamiltonian flow for complex times, which is achieved in the framework of complex eigenvalues (resonances).  
\smallskip
Our last Hypothesis concerns the non-resonance condition relative to Floquet exponents (see [Br]), which is required to achieve
Birkhoff normal form, namely
$$r=d \ \hbox{and} \ \forall k_1,\cdots,k_r \in {\bf Z}: \ \sum_{j=1}^{r} k_j\mu_j\in 2i\pi{\bf Z}  \Longrightarrow  
\sum_{j=1}^{r} k_j \mu_j = 0\leqno(1.7)$$
For instance, when $n=2$ and $\mu_1 = i\omega_1$, it takes the form
$k_1i\omega_1 \in 2i\pi {\bf Z}$, iff $k_1\omega_1=0$, i.e. the rotation number $\omega_1$ is irrational.
We need also the strong non-resonance condition on Floquet exponents: 
$$r=d \ \hbox{and} \ \forall k_1,\cdots,k_r\in{\bf Z}: \ \sum_{j=1}^{r} k_j \mu_j \in 2i\pi {\bf Z}
\Longrightarrow k_j=0, j = 1,\cdots,d\leqno(1.8) $$
\noindent {\it b) Exemples}:

1) The Model Hamiltonian 
$$H_{\mod}(hD_t,x,hD_x;h)=-hD_t+\Sum_{j=1}^d\mu_jQ_j^w(x,hD_x)\leqno(1.9)$$ 
$Q_j^w(x,hD_x)={1\over2}(x_jhD_{x_j}+hD_{x_j}x_j)$, with Periodic Boundary Conditions on ${\bf S}^1\times{\bf R}^d$,
serves as a guide-line throughout this work. Here $x$ may denote complex variables, in some Bargmann representation
of the Hamiltonian.

2) A physical example is given by
$H(y,hD_y)=-h^2\Delta_y+|y|^{-1}+ay_1$ on ${\bf R}^n$ (repulsive Coulomb potential perturbed by Stark effect) near an energy level
$E>2/\sqrt a$. It can be generalized in the case of 3 bumps of potential (``Monkey Saddle''), see [Sj3]. 

3) The geodesic flow on the one-sheeted hyperboloid in ${\bf R}^3$ has an (unstable) periodic orbit of hyperbolic type 
(Poincar\'e example). This example generalises [Chr,App.C] to a surface of revolution in ${\bf R}^4$, involving one pair of real-hyperbolic
elements and two pairs of elliptic elements. Our method easily carries to the case when $H(y,hD_y)$ 
is the geodesic flow on such manifolds. 
\smallskip
Before stating our result we recall the index of a symplectic arc (Gelfand-Lidskiy, or Conley-Zehnder index), which 
appears in the quantization condition when elliptic elements occur.
Standard Maslov index is associated with a differentiable loop in $\Sp(2n;{\bf R})$, while 
Conley-Zehnder index (in the formulation of [SaZe])
is defined for a differentiable path $\Psi:[0,T]\to\Sp(2n;{\bf R})$ such that 
$\Psi(0)=\id$ and $\det(\id-\Psi(T'))\neq0$ 
for some $T'\in]0,T[$. In the present case, let
$Y(t)$ solve the variational system along $\gamma_E$, i.e. $\dot Y(t)=JH''(\Phi^t(\rho))Y(t)$, $Y(0)=\id$, 
where we recall $\Phi^t(\rho)$ is the flow of $X_{H_0}$ issued from $\rho\in\gamma_E$ as in (1.2). We define $\Psi(t)=d{\cal P}_E(t)$ 
as the co-restriction of $Y(t)$ to Poincar\'e sections, i.e. $\Psi:[0,T]\to\Sp(2d;{\bf R})$; here $T$ is taken to be
the period of $\gamma_E$. Then, Conley-Zehnder index can be interpreted as the mean winding number of the eigenvalues of the first kind,
and is computed here most easily using Birkhoff normal form along $\gamma_E$. 

We are concerned with semi-classical {\it resonances} of $H$ near $E_0=0$, in the framework of ``complex scaling''
theory and its extensions [ReSi], [HeSj], i.e. the discrete spectrum of some suitable analytic continuation of $H$
as a closed, Fredholm, but non-selfadjoint operator.
Our main result, the generalized Bohr-Sommerfeld quantization condition, can be formulated as follows:
\medskip
\noindent{\bf Theorem}: {\it Under the hypotheses above, let (after re-ordering) $\mu_j=i\omega_j$, $j=1,\cdots,\ell$,
$\omega_j>0$ be the set of elliptic Floquet exponents for $H_0$. Recall $H_1$ from (1), and let $H_1(x(t),\xi(t))\,dt$ 
the sub-principal 1-form. We define the {\it semi-classical} action along $\gamma_E$, mod ${\cal O}(h^2)$, by
${\cal S}(E;h)=S_0(E)+hS_1(E)+{\cal O}(h^2)$, with
$$\leqalignno{
&S_0(E)={1\over2\pi h}\int_{\gamma_E}\xi\,dx&(1.10)\cr
&S_1(E)=-{1\over2\pi }\int_0^{T(E)} H_1(x(t),\xi(t))\,dt+{1\over2\pi}(\omega_1(E)+\cdots+\omega_\ell(E))+{g_\ell\over2}&(1.11)\cr
}$$
where $\omega_1(E),\cdots,\omega_\ell(E)$ are the actions along (complex) arcs on a Poincar\'e section, 
$\omega_j(E)=\omega_j+{\cal O}(E)$,
and $g_\ell\in{\bf Z}$ Cohnley-Zehnder index of $\gamma_E$.
Then the resonances of $H$ near 0 are given (at first order in $h$) by generalized Bohr-Sommerfeld (BS) quantization condition 
$${1\over2\pi h}{\cal S}(E;h)+{1\over2i\pi h}\Sum_{j=1}^d k_j\mu_j(E)=m\in{\bf Z}, \ k=(k_1,\cdots,k_d)\in{\bf Z}^d\leqno(1.12)$$
provided $|m|h\leq \e_0$, $|k|h\leq h^\delta$, with $0<\delta<1$. }
\smallskip
In the elliptic case, a similar theorem (for real spectrum) was obtained in [Ba], [BaLa], and [Ra];
in the real hyperbolic case, in [G\'eSj] for $|\im E|={\cal O}(h)$, and [Sj4] in dimension 2 with $|\im E|={\cal O}(h^\delta)$
or even $|\im E|$ small enough independently of $h$, but selecting a single Floquet parameter in the semi-classical Floquet decomposition
of $H$ near $\gamma_0$,
i.e. few ``longitudinal'' or ``principal'' quantum numbers $m\in{\bf Z}$.
For related results about trace formulas or concentration
of eigenvalues, see [Vo], [SjZw],
[NoSjZw], [Chr]. For the wave equation outside convex obstacles, see
[Ik], [G\'e]. 
\medskip
\noindent {\bf II. Outline of proof}.
\smallskip
The main object to be constructed is the semi-classical {\it monodromy operator} $M^*(E)$, a $h$-FIO quantizing Floquet operator
associated with the periodic orbit. 
\smallskip
\noindent {\it 1) Birkhoff normal form (BNF)}
\smallskip 
We start to find suitable coordinates near $\overline\gamma$. When $\re\mu_j>0$ for all $1\leq j\leq d$, the stable/unstable
manifold theorem guarantees the existence of involutive manifolds $\Gamma_\pm$ in a neighborhood of $\gamma_0$ with
$$T_\rho\Gamma_++T_\rho\Gamma_-+T_\rho\overline\gamma=T^*{\bf R}^n, \quad \rho\in\gamma_0\leqno(2.1)$$ 
There are (real) symplectic coordinates
$(t,\tau,x,\xi)$ such that
$\xi=0$, $d\xi\neq0$ on $\Gamma_+$, $x=0$, $dx\neq0$ on $\Gamma_-$, and $(t,\tau)$ parametrize $\overline\gamma$.
Write here $H$ instead of $H_0$. Intersecting with the energy surfaces $H^{-1}(E)$ gives the foliation 
$$T_\rho\Gamma_+(E)+T_\rho\Gamma_-(E)+T_\rho\gamma(E)=T_\rho^*H^{-1}(E), \quad \rho\in\gamma(E)\leqno(2.2)$$ 
and $\Gamma_\pm(E)$ are Lagrangian
submanifolds. In these coordinates 
$H(y,\eta)=f(\tau)+\langle B(t,\tau,x,\xi)x,\xi\rangle$.
Here $f$ parametrizes the energy parameter
$f(\tau)=E$, and is related with the period $T(E)$ of $\gamma(E)$ by 
$f'(\tau)={2\pi\over T\circ f(\tau)}$. Performing a first canonical transformation gives $B(t,\tau,x,\xi)=B_0(\tau)+{\cal O}(|\tau|,|x,\xi|)$,
where the eigenvalues of $B_0(\tau)$ are Floquet exponents for ${\cal P}_E$ with positive real part. 
When $\re\mu_k=0$ for some $k$, (2.1) and (2.2) still hold provided we take complex variables. 
In both cases however, under the non resonance conditions (1.7), (1.8) BNF holds in the classical sense [Bry] as well as in the
semi-classical sense [GuPa] and takes, modulo a small remainder term,
operator $H^w(y,hD_y;h)$ to a polynomial in $hD_t$ and $Q_j^w(x,hD_x)$,
$Q_j(x,\xi)$ being one of the quadratic polynomials above. In particular, the principal part of $H^w(y,hD_y;h)$ in 
in BNF is given by (1.9) in some suitable Bargmann (still formal) representation, provided a reparametrization of energy.
\smallskip
\noindent {\it 2) Microlocalisation in the complex domain}
\smallskip 
As trying to construct quasi-modes for $H^w(y,hD_y;h)$ microlocalized near the stable/unstable manifolds $\Gamma_\pm$
one meets the difficulty, already observed in [Du], that because of hyperbolic elements,
there is no smooth $X_{H_0}$-invariant density on $\Gamma_\pm$. On the other hand, elliptic elements are responsible
for caustics in the time-evolution (or Cauchy problem). These difficulties naturally disappear 
in the framework of resonances, provided we are working within the framework of $h$-FIO's with complex phase of positive imaginary part,
as in [MeSj], [Sj], see also [M]. Resonances here are considered from the point of vue of analytic dilations
and Lagrangian deformations; taking into account that there exists an {\it escape function} (that grows along the flow of $X_{H_0}$)
which implies kind of a ``virial condition'' outside the trapped set
$\gamma_0$, the 
most relevant region of phase-space for such deformations is a neighborhood of $\gamma_0$.
Here we make a complex scaling of the form $(x,\xi)\mapsto(e^{i\theta}x,e^{-i\theta}\xi)$, followed also by a small
deformation in the $(t,\tau)$ variables. Our main tool is
FBI transformation (metaplectic FIO with complex phase) which takes the form, in coordinates 
$(s,y;t,x)\in T^*{\bf R}^n\times T^*{\bf C}^n$  adapted to $\Gamma_\pm$ as in BNF
$$T_0u(x,h)=\int e^{i\varphi_0(t,s;x,y)/h}u(s,y)\,ds\,dy, \ u\in L^2({\bf R}^n)$$ 
where $\varphi_0(t,s;x,y)=\varphi_1(t,s)+\varphi_2(x,y)$,
$\varphi_1(t,s)={i\over2}(t-s)^2$, $\varphi_2(x,y)={i\over2}\bigl[(x-y)^2-{1\over2}x^2\bigr]$.
The corresponding canonical transformation is $\kappa_0=(\kappa_1,\kappa_2)$, with
$\kappa_1:(s,-\partial_s\varphi_1)\mapsto(t,\partial_t\varphi_1)$, $\kappa_2:(y,-\partial_y\varphi_2)\mapsto(x,\partial_x\varphi_2)$,
and the corresponding pl.s.h. weight $\Phi_0=\Phi_1+\Phi_2$, with 
$$\Phi_1(t)=\sup_{s\in{\bf R}}(-\im \varphi_1(t,s))=(\im t)^2/2, \ \Phi_2(x)=\sup_{y\in{\bf R}^d}(-\im \varphi_2(x,y))=|x|^2/4$$ 
In a very small neighborhood of $\gamma_0$, whose size
will eventually depend on $h$, corresponding to $\theta=-\pi/4$, and that we call the ``phase of inflation'', it turns out
that $H^w(y,hD_y;h)$ takes the simple form above whose principal term is given in (1.9), and the corresponding weight 
$\widetilde\Phi(t,x)$ is just
$\Phi_0(t,x)$. Otherwise we take $\theta$ small enough in a somewhat larger neighborhood of $\gamma_0$, which we call
the ``linear phase''. These weights are
deformed continuously in phase-space, depending on the escape function, and patched together in overlapping regions,
so to define a globally pl.s.h. function in complex $x$-space. They also
define the contour integral for writing a $h$-FIO in the complex domain; when the weight is quadratic, and in the particular
case of a $h$-PDO with $C^\infty$ amplitude $a$
$$Av(z,h)=\int _{\Gamma(z)}e^{i(z-y)\eta/h}a((z+y)/2,\theta,h)v(y)\, dy\wedge d\eta$$
we have $\Gamma(z)=\{\eta={2\over i}{\partial\Phi\over\partial z}({z+y\over2}); y\in{\bf C}^n\}$. 
Where $a$ is not analytic, due in particular to BNF, $a$ denotes an
almost analytic extension. When the weight is not quadratic, they still define ``good contours'' in the sense of [Sj].
\smallskip
\noindent {\it 3) Poisson operator and its adjoint}
\smallskip 
We look for $K(t,E):L^2({\bf R}^d)\mapsto L^2({\bf R}^d)$, microlocalized near $\Gamma_+(E)$, 
such that 
$$H(hD_t,x,hD_x;h)K(t,E)=0, \ K(0,E)=\id$$
In the phase of inflation, this is an OIF $H_{\Phi_2}({\bf C}^d)\to H_{\Phi_2^t}({\bf C}^d)$
$$K(t,E)v(x;h)=\int\int e^{i(S(t,x,\eta)-y\eta)/h}a(t,x,\eta;E,h)v(y)\,dy\wedge d\eta$$
where $\Phi_2^t(x)=\Phi_2\circ\kappa_t$ (where $\kappa_t$ denotes the flow of $X_{H_0}$ in complex coordinates, when
restricted to $(x,\xi)$ variables),
and the integration is carried over a suitable contour as above. 
We solve eikonal and transport equations for $S$, with initial condition $S(0,x,E)=x\eta$, and find for instance
$S=S_0+\widetilde S$, where
$$S_0(t,x;\eta)=-Et+\Sum_{j=1}^dx_j\eta_je^{\mu_jt}$$
is the phase for the model Hamiltonian, and the correction $\widetilde S(t,x,E)$ has the asymptotic form
$$\widetilde S(t,x,E)=-t\Sum_{k\geq2}s_kE^k+\Sum_{\alpha,l,m}c_{\alpha,l,m}E^\alpha Q^\ell(x,\eta) e^{\langle m,\mu\rangle t}$$
with $|\ell|+\alpha\geq\sup(|m|,2)$. So $K(t,E)$ is in BNF, i.e. its kernel depends on $(x,\eta)$ only through $Q_j(x,\eta)$.
Its adjoint $K^*(E)=\int_{\gamma^*(E)} K^*(t,E)\,dt$ is continuous $\int^\oplus H_{\Phi_2^t}({\bf C}^d)\to H_{\Phi_2}({\bf C}^d)$
provided $\gamma^*(E)=\{|\im t|=\e \}$, $0<\e <2|E+\Sum_{k\geq2}s_kE^k|$. This fixes the relative sizes between $\tau$
and $|x,\xi|^2$.
In the model case we have simply
$$K(t,E)v(x,h)=(2\pi h)^{-d}\int\int e^{i(-Et+\sum x_j\eta_je^{\mu_j t}-y\eta)/h}e^{(\sum \mu_j)t/2}v(y)\,dy\,d\eta$$
\smallskip
\noindent {\it 4) Normalization}.
\smallskip
We proceed first formally, i.e. with operators in the real domain.
We use the ``flux norm'' of [SjZw] to normalize Poisson operator.
Let $\chi\in C^\infty({\bf R})$, be equal to 0 near 0, 1 near $[2\pi,\infty[$.
Pseudo-differential calculus shows that there is a $h$-PDO $B(E)=B^w(x,hD_x;E)$ such that 
$L(t,E)=K(t,E)B(E)$ satisfies 
$$\bigl({i\over h}[H,\chi(t)]L(t,E)v|L(t,E)v\bigr)=\bigl(v|v\bigr)\leqno(2.3)$$
that is $\int dt\,L(t,E)^*{i\over h}[H,\chi(t)]L(t,E)=\id_{L^2({\bf R}^d)}$.
In the other way, we need also consider operators like $\int L(s,E)L^*(E)\,dE$, or $\int dE\,L^*(E)L(s,E)$. In the model case
$$\int dE\,L(s,E)L^*(E)u(s,x)=\int\int e^{i(-E(s-t)+(x-y')\eta')/h}e^{\widetilde\mu (s-t)/2}u(t,y'e^{\mu(s-t)})\,dy'\,d\eta'\,dt\,dE=u(s,x)$$
by asymptotic Fourier inversion formula, since $e^{\widetilde\mu (s-t)/2}u(t,y'e^{\mu(s-t)})$ is independent of $(E,\eta')$.
In the general case, we only have $\int dE\,L(\cdot,E)L^*(E)=\id_{L^2({\bf R}^n)}+{\cal O}(h)$, and similarly
$\int dE\,L^*(E)L(s,E)=\id_{L^2({\bf R}^n_s)}+{\cal O}(h)$, denoting by ${\bf R}^n_s$ the section $\{s\}\times{\bf R}^d$
of ${\bf R}^n$ (in BNF coordinates). 
By Pseudo-differential calculus there exists $P(s,E)=\id_{L^2({\bf R}^n_s)}+{\cal O}(h)$ such that
$$\int dE\,L(\cdot,E)P(\cdot,E)L^*(E)=\id_{L^2({\bf R}^n)}, \ \int dE\,L^*(E)L(s,E)P(s,E)=\id_{L^2({\bf R}^n_s)}\leqno(2.4)$$
These (formal) computations can be carried out in the framework of FIO's in the complex domain. A similar situation was met in [BdMSj]
when considering Bergman and Szeg\"o projectors.
\smallskip
\noindent {\it 5) The monodromy operator}.
\smallskip
We set $K_0(t,E)=K(t,E)$ where $K(t,E)$ is Poisson operator with Cauchy data at $t=0$, and $L_0(t,E)=K_0(t,E)B(E)$; we set similarly 
$L_{2\pi}(t,E)=K_0(t-2\pi,E)B(E)$. The monodromy operator (or semi-classical Poincar\'e map) is defined by 
$$M^*(E)=L_{2\pi}^*(E){i\over h}[H,\chi]L_0(\cdot,E)\leqno(2.5)$$ 
as an operator on $L^2({\bf R}^d)$.
Actually, as a function de $\chi$, $M^*(E)$ follows a ``0-1 law'': it is 0 if $\supp \chi\subset]0,2\pi[$, and unitary if 
$\chi$ equals 0 near 0, and 1 near $2\pi$. For the model case one has 
$M^*(E)v(x)=e^{-2i\pi E/h}e^{\pi \mu}v(xe^{2\pi \mu})$ when $\int\chi'(t)\, dt=1$.
We check the unitarity of $M^*(E)$ as follows: by the first equality (2.4)
$$\eqalign{
M^*(E)&M(E)=L_{2\pi}^*(E){i\over h}[H,\chi]L_0(\cdot,E)L_0^*(E){i\over h}[H,\chi]L_{2\pi}(\cdot,E)\cr
&=L_{2\pi}^*(E)\bigl(\int dE'\, L_0(\cdot,E')P_0(\cdot,E')L_0^*(E')\bigr){i\over h}[H,\chi]L_0(\cdot,E)L_0^*(E)\cr
&\bigl(\int dE''\, L_{2\pi}(\cdot,E'')P_{2\pi}(\cdot,E'')L_{2\pi}^*(E'')\bigr){i\over h}[H,\chi]L_{2\pi}(\cdot,E)\cr
}$$
that is
$$\eqalign{
M^*(E)&M(E)=L_{2\pi}^*(E)\int dE'\, L_0(\cdot,E')P_0(\cdot,E')\bigl(L_0^*(E'){i\over h}[H,\chi]L_0(\cdot,E)\bigr)L_0^*(E)\cr
&\int dE''\, L_{2\pi}(\cdot,E'')P_{2\pi}(\cdot,E'')\bigl(L_{2\pi}^*(E''){i\over h}[H,\chi]L_{2\pi}(\cdot,E)\bigr)\cr
}$$
Expanding the kernels by stationary phase, we 
take in account (2.3) to estimate the contribution of 
$L_0^*(E'){i\over h}[H,\chi]L_0(\cdot,E)$ and $L_{2\pi}^*(E''){i\over h}[H,\chi]L_{2\pi}(\cdot,E)$
and use eventually the second equality (2.4). So $M^*(E)M(E)=\id$, and similarly $M(E)M^*(E)=\id$.
We then check the structure of $M^*(E)$ using BNF, and find that 
$$M^*(E)v(x,h)=\int dt\,\int{\cal M}^*(t,x,z)v(z)\,dz$$
with the kernel ${\cal M}^*(t,x,z)$  of the form
$${\cal M}^*(t,x,z)=\chi'(t)\int e^{i\Psi_2(t,x,z,\eta')/h}m(t,x,z,\eta';h)\,d\eta'$$
which is in BNF, and where the integral is independent of $t$. In fact
$M^*(E)=e^{iR^w(x,hD_x;E,h)/h}$,
where $R$ is $h$-PDO in BNF, self-adjoint for real $E$. 
\smallskip
\noindent {\it 6) End of the proof}.
\smallskip
Reducing the spectral problem for $H^w(x,hD_x;h)$ through a Grushin operator as in [SjZw], [FaLoRo], 
we consider the approximate kernel of $M^*(E)-\id_{L^2({\bf R}^d)}$ for complex values of $E$. 
In particular we know that 
for real $E$, $M^*(E)$ is (microlocally) unitary with absolutely continuous spectrum,
while 1 belongs to its discrete spectrum for some complex values of $E$, which are precisely the resonances. 
This requires first to take suitable analytic extensions with respet to $E$ of Poisson and monodromy operators. 
Since $M^*(E)$ is in BNF (formal) eigenfunctions of $M^*(E)$ are
the homogeneous polynomials $f_\alpha(x)$ (of degree depending on the accuracy of BNF, and that we can take of order $h^{-\delta'}$,
for some $0<\delta'<1$). This gives the ``transverse'' quantum numbers. The longitudinal quantum number $m$ is found by taking also the 
phase of $M^*(E)f_\alpha$ equal to $2k\pi$, $k\in{\bf Z}$, of order also depending on the accuracy of BNF.
Thus the Theorem is proved.
\medskip
\noindent {\it Acknowledgements}: We thank Prof. Alain Chenciner for pointing to us interesting references on the subject, in particular [A].
\medskip
\noindent {\bf References}
\smallskip
\noindent [A] Marie-Claude Arnaud. On the type of certain periodic orbits minimizing the Lagrangian action. Nonlinearity 11, p.143-150, 1998.

\noindent [ArKoNe]  V.Arnold, V.Kozlov, A.Neishtadt. Mathematical aspects of classical and celestial mechanics. Encyclopaedia of Math. Sci.,
Dynamical Systems III, Springer, 2006.

\noindent [Ba] V.Babich. Eigenfunctions concentrated near a closed geodesic [in Russian], Vol.9, Zapiski Nauchnykh Seminarov 
LOMI, Leningrad, 1968. 

\noindent [BLaz] V.M.Babich, V.Lazutkin.  Eigenfunctions concentrated near a closed geodesic. Topics in Math. Phys., Vol.2, M.Birman, ed. 
Consultants' Bureau, New York, 1968, p.9-18

\noindent [BdMSj L.Boutet de Monvel, J.Sj\"ostrand. Sur la singularit\'e des noyaux de Bargman et de Szeg\"o. Ast\'erisque 34-35, p.123-164, 1976.

\noindent [Br] A.D.Bryuno. {\bf 1}. The normal form of an Hamiltonian system. Russian Math. Surveys 43:1, p.25-66, 1988.
{\bf 2}. Normalization of a Hamiltonian system near an invariant cycle or torus. Russian Math. Surveys 44:2, p.53-89, 1991.

\noindent [Chr] H.Christianson. {\bf 1}. Semiclassical non-concentration near hyperbolic orbits. J.Funct.Anal. 246, p.145-195, 2007.
{\bf 2}. Quantum monodromy and nonconcentration near a closed semi-hyperbolic orbit. Trans. Amer. Math. Soc. 363, No.7, p.3373–3438, 2011. 

\noindent [Du] J.J.Duistermaat. Oscillatory integrals, Lagrangian immersions and unfolding of singularities. Comm. Pure Appl. Math.
27, p.207-281, 1974.

\noindent [FaLoRo] H.Fadhlaoui, H.Louati, M.Rouleux. Hyperbolic Hamiltonian flows and the semiclassical Poincar\'e map. 
Proceedings ``Days of Diffraction 2013'', Saint-Petersburg, IEEE 10.1109/ DD.2013. 6712803, p.53-58.

\noindent [HeSj] B.Helffer, J.Sj\"ostrand. R\'esonances en limite semi-classique. M\'emoires S.M.F. 114(3), 1986.

\noindent [G\'e] C.G\'erard. Asymptotique des p\^oles de la matrice de scattering pour 2 obstacles strictement convexes. M\'emoire Soc.
Math. France, S\'er.2 (31), p.1-146, 1988.

\noindent [G\'eSj] C.G\'erard, J.Sj\"ostrand. {\bf 1}. Semiclassical resonances generated by a closed trajectory of hyperbolic type.
Comm. Math. Phys. 108, p.391-421, 1987. {\bf 2}. Resonances en limite semiclassique et exposants de Lyapunov.
Comm. Math. Phys. 116, p.193-213, 1988.

\noindent [GuWe] V.Guillemin, A.Weinstein. Eigenvalues associated with a closed geodesic. Bull. AMS 82, p.92-94, 1976.

\noindent V.Guillemin, T.Paul. Some remarks about semiclassical trace invariants and quantum normal forms. 
Comm. Math. Phys. 294 (2010), no. 1, 1–19.

\noindent [IaSj] A.Iantchenko, J.Sj\"ostrand. Birkhoff normal forms for Fourier integral operators II. American
J. of Math., 124(4), p.817-850, 2002.

\noindent [I] M.Ikawa. On the existence of poles of the scattering matrix for several convex bodies. Proc. Japan Acad. Ser.A Math. Sc., 64,
p.91-93, 1988.

\noindent [KaKe] N.Kaidi, Ph.Kerdelhue. Forme normale de Birkhoff et r\'esonances. Asympt. Analysis 23, p.1-21, 2000.

\noindent [LasSj] B.Lascar, J.Sj\"ostrand. Equation de Schr\"odinger et propagation des singularit\'es 
pour des OPD à caract\'eristiques
de multiplicit\' variable, I (Ast\'erisque No.95, 1982), and II, Comm. Part. Diff. Eq., 1985.

\noindent [M] V.P.Maslov. Th\'eorie des Perturbations et M\'ethodes Asymptotiques. Dunod, Paris, 1972.

\noindent [MeSj] A.Melin, J.Sj\"ostrand. {\bf 1}. FIO's with complex valued phase functions. Springer Lect. Notes in Math.459, p.120-223.
{\bf 2}. Bohr-Sommerfeld quantization condition for non-self-adjoint operators in dimension 2.

\noindent [MoZh] J.Moser, E.Zehnder. Notes on Dynamical systems. American Math. Soc., Courant Inst. Math. Sci. Vol.12, 2005.

\noindent [NoSjZw] S.Nonnenmacher, J.Sj\"ostrand, M.Zworski. From Open Quantum Systems to Open Quantum maps.
Comm. Math. Phys. 304, p.1-48, 2011

\noindent [Ra] J.V.Ralston. On the construction of quasi-modes associated with periodic orbits. Comm. Math. Phys. 51(3) p.219-242, 1976.

\noindent [ReSi] M.Reed, B.Simon. Methods of Modern Math.Phys. IV Analysis of Operators. Acad. Press, 1978. 

\noindent [SaZe] D.Salamon, E.Zehnder. Morse Theory for Periodic Solutions of Hamiltonian Systems
and the Maslov Index. Comm. on Pure and Appl. Math., Vol. XLV, p.1303-1360, 1992.

\noindent [Sj] J.Sj\"ostrand. {\bf 1}. Singularites analytiques microlocales. Asterique No 95, 1982. 
{\bf 2}. Semi-classical resonances generated by a non-degenerate critical point, 
{\it in} Lect.Notes in Math. Vol.1256,
Springer, p.402-429. 
{\bf 3}. Geometric bounds on the density of resonances for semiclassical problems. Duke Math. J. Vol.60, No.1, p.1-57, 1990.
{\bf 4}. Resonances associated to a closed hyperbolic trajectory in dimension 2. Asympt. Analysis 36, p.93-113, 2003.

\noindent [SjZw] J. Sj\"ostrand and M. Zworski. Quantum monodromy and semi-classical trace formulae, J. Math. Pure Appl.
81(2002), 1-33. Erratum: http://math.berkeley.edu/~zworski/qmr.pdf

\noindent [Vo] A.Voros. Semi-classical approximations. Ann. Inst. H.Poincar\'e, 24, p.31-90, 1976.
{\bf 2}. Unstable periodic orbits and semiclassical quantization. J.Phys. A(21), p.685-692, 1988.

\bye

%% file: Definitions.tex
\magnification=1100

\hsize 17truecm
\vsize 23truecm

\font\twelvec=msbm10 at 10pt
\font\sevenc=msbm10 at 7pt
\font\fivec=msbm10 at 5pt

\newfam\co
\textfont\co=\twelvec
\scriptfont\co=\sevenc
\scriptscriptfont\co=\fivec

\def\const{\mathop{\rm const.}\nolimits}
\def\det{\mathop{\rm det}\nolimits}

\def\exp{\mathop{\rm exp}\nolimits}

\def\id{\mathop{\rm Id}\nolimits}
\def\im{\mathop{\rm Im}\nolimits}

\def\e{\mathop{\rm e}\nolimits}

\def\lim{\mathop{\rm lim}\nolimits}

\def\mod{\mathop{\rm mod}\nolimits}

\def\re{\mathop{\rm Re}\nolimits}

\def\Sp{\mathop{\rm Sp}\nolimits}

\def\supp{\mathop{\rm supp}\nolimits}

\def\sup{\mathop{\rm sup}\nolimits}

\def\Sum{\displaystyle\sum}
\def\e{\mathop{\rm \varepsilon}\nolimits}

\baselineskip 15pt